\documentclass[aps,prapplied,reprint,groupedaddress,amsmath]{revtex4-2}
\usepackage{graphicx}
\usepackage{epstopdf}

\begin{document}


\title{Angular distribution of electron emission from ultrafast nanotip sources}


\author{Jonathan M. Geller}
\author{Michael J. Faulkner}
\author{Iona K. Binnie}
\author{Catherine Kealhofer}
\email[]{ck12@williams.edu}
\affiliation{Williams College Department of Physics, Williamstown, MA 01267}


\date{\today}
\begin{abstract}
We investigate the angular distribution of ultrafast laser-induced electron emission from a tungsten nanotip in the multiphoton regime. A theoretical model allows precise determination of the relative contribution of different electron emission mechanisms, revealing connections between emission mechanism and the angular distribution of emitted electrons. We infer a continuous map of the work function across the surface of the tip, which in combination with the model can be used to predict values including the number of electrons per pulse and the angular divergence of the resulting beam as a function of laser power and tip voltage for (310)-oriented tungsten nanotips.  The model is straightforward to implement and can be used to optimize the performance of instruments using ultrafast nanotip electron sources.

\end{abstract}

\maketitle

\section{Introduction\label{Sec:Intro}}
Pulsed electron sources are at the heart of many novel electron microscopy techniques.  They enable time-resolved measurements in dynamical transmission microscopy \cite{Doemer2003, Kim2008} and ultrafast electron diffraction and microscopy \cite{Siwick2003, Flannigan2012}. 
Techniques like time-resolved cathodoluminescence can give access to excited state lifetimes and carrier mobility in samples with nanometer spatial resolution \cite{Kim2021, Meuret2021}.  Photon-induced near-field electron microscopy \cite{Barwick2009, GarciadeAbajo2008} provides a mechanism for imaging optical near fields in nanostructures  and has given rise to myriad applications ranging from manipulating the quantum state of free electrons \cite{Feist2015} to proposals for efficiently generating highly non-classical states of light \cite{BenHayun2022}.  In the domain of biological imaging, pulsed electron sources could be used to reduce specimen damage via multipass electron microscopy \cite{Koppell2019, Reynolds2023}.  

Nanotip electron sources, which are tapered wires ending in a hemisphere with a $\sim 5$--100~nm radius of curvature, have effective source sizes on the scale of 1--2~nm \cite{Tsujino2022}, resulting in very low emittance, high brightness beams with large transverse coherence lengths.  These properties can be exploited for high resolution electron microscopy \cite{Crewe1968a} and phase-sensitive techniques such as electron holography and Lorentz microscopy \cite{Houdellier2018, Feist2017}.  Nanotips also enable electron probe sizes that are substantially smaller than an optical wavelength, a crucial property for techniques like photon-induced near-field electron microscopy.

In a conventional cold field emission electron gun, a strong electric field at the surface of a nanotip induces electrons to tunnel into the vacuum, creating a continuous electron beam.  By contrast, ultrafast nanotip sources use a weaker bias field in conjunction with short laser pulses to trigger electron emission.  At comparatively low laser power, the high peak intensity of an ultrafast pulse, increased by local optical field enhancement, can lead to emission via mechanisms including photo-assisted field emission and single- and multi-photon over-the-barrier emission. In these emission processes, the intensity profile of the exciting laser pulse is mapped onto the temporal profile of the emitted electron pulse \cite{Ropers2007}.  For more intense, few-cycle pulses, the strong field emission regime may be reached, which results in sub-optical cycle electron emission \cite{Schenk2010}.

Alongside emittance and brightness, a key figure of merit is the angular divergence of the source.    
Increased angular divergence leads to longer pulses, since angular coordinate is correlated with arrival time at a sample \cite{Weninger2012}.  For simple electron optical systems with spherical aberrations, it also leads to larger focused spot sizes.   Although apertures in the optical system can be used to limit the angular divergence, this means that the number of electrons in the pulse in the electron gun region must be significantly larger than the number of electrons available for an experiment.  Coulomb repulsion in the vicinity of the nanotip degrades achievable pulse parameters at the gun exit, including emittance, transverse coherence, longitudinal energy spread, and pulse duration \cite{Cook2016, Meier2022}.

The shape of the angular distribution of emission from a nanotip is dependent on a number of factors.  Chief among these is the emission mechanism.  DC field emission typically gives rise to a narrow emission pattern, but under ultrafast illumination, the emission pattern may be qualitatively different for different regimes. As a consequence, although the number of electrons per pulse may be increased by increasing laser intensity or bias field, such changes may negatively impact the emission pattern and electron energy distribution.  An accurate model for the initial distribution of electron coordinates is crucial for developing realistic electron optics simulations and optimizing experiment design.

With this goal, we develop a general model that can be used to predict angular divergence and establish limits on the number of electrons per pulse in different multiphoton emission regimes.  We experimentally characterize the angular distribution of electrons in laser-induced electron emission from a tungsten nanotip for two- and three-photon photo-assisted field emission and over-the-barrier emission to find empirical values for key parameters in the model.  These results provide necessary inputs for nanotip electron gun optimization and pave the way for improved spatio-temporal resolution in experiments that require ultrafast electron pulses.

\section{Theoretical model\label{Sec:Background}}
\subsection{Model for multiphoton emission current density} \label{subsec:MultiphotonTheory}

In a strong applied field, electrons inside the metal can tunnel into the vacuum.  The potential barrier at the surface has an approximately triangular shape, modified by the image potential of the electron as it leaves the metal.  For an electron of a given energy inside the metal, the probability of transmission through the barrier $D$ was calculated by Murphy and Good \cite{Murphy1956} as a function of the electron's kinetic energy normal to the surface, $W$.  The transmission coefficient depends also on the shape of the barrier, which is determined by the applied electric field $F$, the work function $\phi$, and the kinetic energy of an electron at the Fermi level, $\mu$:
\begin{align}
&D(W, F) =  \nonumber \\
  &\left\{1+\exp \left(\frac{4\sqrt{2}}{3}\left[\frac{(4\pi\epsilon_0)^3\hbar^4 F}{m^2e^5}\right]^{-\frac{1}{4}}y^{-3/2} v(y)\right) \right\}^{-1},
	\label{eq:TransmissionFunction}
\end{align}
	where $y = \sqrt{\frac{e^3 F}{4\pi\epsilon_0}} \frac{1}{|W-\phi-\mu|}$, and $v(y)$ is the elliptic integral
	\[
	v(y) = -\frac{3 i}{4\sqrt{2}} \int_{1-\sqrt{1-y^2}}^{1+\sqrt{1-y^2}}\sqrt{\rho-2+y^2\rho^{-1}}d\rho.
\]
	This expression is valid when the normal energy $W$ is below the limiting value $W_L \equiv \mu + \phi -\frac{1}{\sqrt{2}} \sqrt{\frac{e^3 F}{4\pi\epsilon_0}}$; above $W_L$, the transmission function is approximately one.

To find the emitted current density, the barrier transmission function can be combined with a supply function, $N(W, T)$, providing the rate that electrons of a given normal energy $W$ are incident on the barrier at temperature $T$.  The emitted current density in a particular direction is thus
	\begin{equation}
	J = e\int_0^\infty N(W, T)\, D(W, F)\; dW,
	\label{eq:SupplyAndTransmission}
	\end{equation}
where $e$ is the elementary charge.

Following reference \cite{Girardeau-Montaut1995}, we separate the emitted current into components corresponding to the absorption of different numbers of photons:   
\begin{equation}
J = J_{\text{DC}} + \sum_{n>0} J_n.
\label{eq:MultiphotonBreakdown}
\end{equation}
DC emission processes (such as field emission and thermionic emission) correspond to absorbing zero photons. 

\begin{figure}%
\includegraphics[width=\columnwidth]{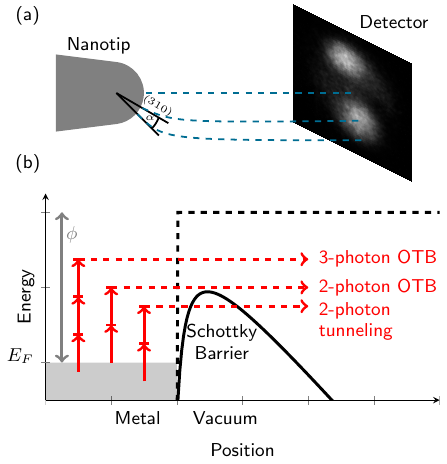} 
\caption{\label{fig:EmissionProcesses} (a) Geometry for measuring emission pattern.  $\alpha$ refers to the angle of emission relative to the (310) direction.  (b) Emission processes: three-photon over-the-barrier emission, two-photon over-the-barrier emission and tunneling (photo-assisted field emission).}%
\end{figure}

We need to define a modified supply function for each photon order, describing electrons that have been excited by a specific number of photons from the original distribution.  At this point, we make several simplifying assumptions \cite{Girardeau-Montaut1995}: (a) electrons in the metal are modeled as a free electron gas, (b) the probability of photon absorption is independent of the electron's initial state, and (c) photon absorption causes the normal energy to increase by $n$ times the photon energy. Approximation (a) is also used in the conventional derivation of the Fowler-Nordheim equation for field emission.  Assumptions (b) and (c) are consistent with studies of cw single-photon photo-assisted field emission \cite{Venus1983}, which showed that emission is dominated by surface-effect photoemission in the commonly-used geometry where the laser is incident at right angles to the tip axis and polarized parallel to the tip axis.  Since very few bulk transitions can simultaneously satisfy energy and momentum conservation, surface effect photoemission, where the electron is effectively excited to a virtual state before tunneling, tends to dominate and essentially replicates the original supply function, shifted by the photon energy \cite{Bagchi1974} (or in the case of multiphoton photoemission, integer multiples of the photon energy).  This model also neglects electron-electron scattering, which will reshape the excited electron distribution and might become noticeable for longer pulses with lengths of order hundreds of femtoseconds \cite{Wu2008}.  
The relevant temperature is the temperature of the tip immediately before the laser pulse arrives. With these assumptions, the $n$-photon component of the current density is
\begin{equation}
J_n = a_n I^n(t) e\int_{n\hbar\omega}^\infty N_n (W, T)\, D(W, F)\; dW.
\label{eq:MultiphotonComponent}
\end{equation}
The coefficient $a_n$ is related to the probability of $n$-photon absorption, and $I(t)$ is the laser intensity as function of time.  Using assumptions (a)--(c), we derive the supply function for $n$-photon emission:
\begin{align}
N_n(W, T) =& \;\frac{m k_B T}{2\pi^2 \hbar^3} \frac{\sqrt{W}}{\sqrt{W-n\hbar\omega}}\nonumber \\ 
 &\times \log\left(1-e^{-(W-n\hbar\omega - \mu)/k_B T}\right),
\label{eq:SupplyFunction}
\end{align}  
with $m$ the electron mass, $\hbar \omega$ the photon energy, $k_B$ Boltzmann's constant, and $T$ the temperature of the electrons.
Equation \ref{eq:MultiphotonComponent} can be integrated numerically and combined with equation \ref{eq:MultiphotonBreakdown} to compute the emitted current density.  As the transmission coefficient is defined for energies both above and below the barrier, this theory includes both over-the-barrier processes, where the excited electron's energy exceeds the Schottky barrier, and tunneling processes (also called photo-assisted field emission), where the excited electron tunnels through the surface potential barrier \cite{Neumann1971, Lee1973}, as shown schematically in Fig.~\ref{fig:EmissionProcesses}.  

\subsection{Factors that impact emission pattern}

The emission pattern reflects the variation in current density over the surface of the tip.  Because there is typically a strong bias field at the tip surface, electrons are rapidly accelerated normal to the tip surface, and the angular distribution of emission is largely determined by the current density distribution at the tip surface and weakly affected by the initial transverse momentum of the electrons, which we neglect here.

The current density distribution is determined by several factors.  First, the work function varies across the surface of the tip.  The curved tip surface intersects the crystal structure to expose different surface orientations.  Assuming a hemispherical shape for the apex, the work function at points on the surface of the tip can be viewed as a function of angular coordinates alone.  This map will be approximately independent of the tip radius once the tip radius is much larger than the interatomic spacing.  Second, the electric field is strongest at the apex and decreases slightly as the polar angle $\theta$ increases; how rapidly the field decays depends on the geometry, especially the opening angle of the tip and to a lesser extent the distance from the tip to an extractor electrode.  
Finally, the optical fields near the tip apex are enhanced resulting in spatial variations of the light intensity at the tip surface.  The spatial variation of this optical field enhancement depends on the tip's radius of curvature and opening angle.  Altering the polarization changes the pattern of optical field enhancement and has been used to favor emission from particular locations on the tip \cite{Yanagisawa2009, Yanagisawa2010}.
\begin{figure}%
\includegraphics[width=\columnwidth]{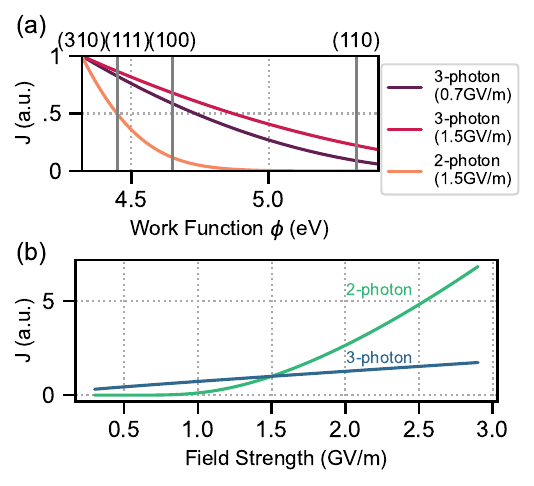}
\caption{\label{fig:WorkFunction_FieldVariation} Variation of theoretical current density with work function and bias field, computed using Eq.~\ref{eq:MultiphotonComponent}, with $\mu= 9.75$~eV for tungsten and a photon energy of $1.575$~eV. (a) Work function dependence.  Curves are normalized to the current density in the (310) direction.  Vertical lines mark the work functions of the (100), (110), and (111) directions (values from \cite{Kawano2022}).  (b) Bias field dependence for $\phi = 4.32$~eV.  The curves are normalized to their value at $F=1.5$~GV/m.}%
\end{figure}

\begin{figure*}%
\includegraphics[width=17.2cm]{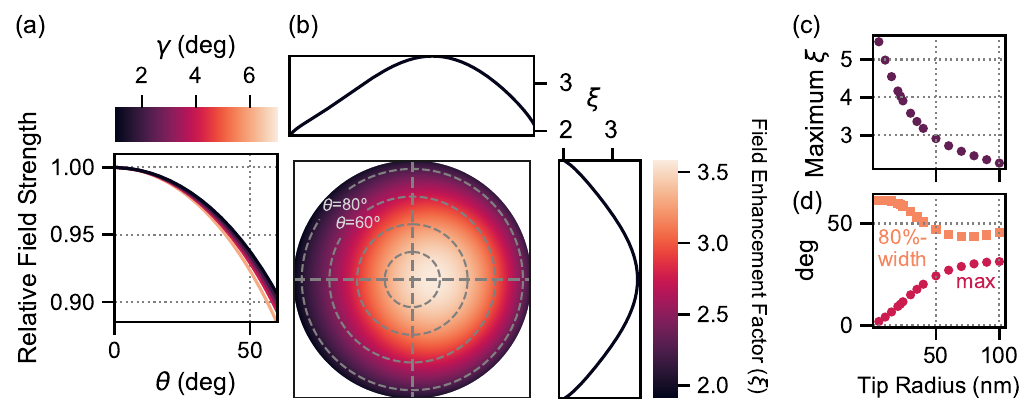}
\caption{\label{fig:FieldEnhancement} (a) Finite element model of the electric field strength relative to the apex electric field, with a distance of 2~cm between the tip and a grounded reference plane.  The cone half angle is $\gamma$.  Field strength is plotted vs.~$\theta$ measured from the tip's axis.  (b) Finite element model of the optical field enhancement $\xi$ at the surface of a tungsten tip of radius 30~nm, with $\gamma = 3.5^\circ$.  The tip is illuminated with an 800~nm plane wave incident from the left and polarized parallel to the axis of the tip.  (c) Peak field enhancement as a function of tip radius.  (d) Angle from the apex at which the maximum field enhancement occurs, and approximate width of the field enhancement region, defined as the half-angle at which the field enhancement has decreased to 80\% of its peak value.}%
\end{figure*}

Figure \ref{fig:WorkFunction_FieldVariation} (a) and (b) plot the work function and field dependence of the calculated current density for different emission regimes: a three-photon-dominated emission regime at lower bias field and higher laser intensity, and a two-photon-dominated regime at higher bias field.  In Fig.~\ref{fig:WorkFunction_FieldVariation} (b), the curves have been normalized to the current density at 1.5~GV/m.  The relative amplitudes of the two-photon and three-photon curves depend on the applied laser intensity.  Note that, for a given laser intensity, due to the larger slope of two-photon emission vs.~field, at sufficiently large field values two-photon emission becomes dominant.  

The surface DC field strength as a function of emission angle relative to the apex is shown in Fig.~\ref{fig:FieldEnhancement}(a) for different cone opening angles.  The field profile was calculated using the finite element method \cite{comsol60} for a conical tip terminated by a spherical end-cap.  It is insensitive to the radius of curvature of the end-cap over the typical range of tip radii.

Figure \ref{fig:FieldEnhancement}(b)--(d) shows the variation of the optical field enhancement over the surface of the tip for some representative geometries.  We define the optical field enhancement factor $\xi$ as the ratio of the amplitude of the electric field component polarized normal to the surface to the amplitude of the incident electromagnetic wave, which is incident from the left side and polarized parallel to the tip axis.  The optical wavelength simulated is 800~nm.  For very small tip radii, the maximum field is near the apex.  As the tip radius increases, the maximum optical field gradually moves towards the side of the tip that faces away from the laser, i.e.~into the geometrical shadow of the tip.

The overall emission pattern is thus determined by the interplay of several effects---the work function variation and the spatial variation of DC and optical fields.  Within a region near the apex of the tip, where variation of DC field strength and optical intensity is limited, localization of emission is primarily due to variation in the work function.  This is especially the case for two-photon emission, because, as shown in Fig.~\ref{fig:WorkFunction_FieldVariation}(a), the sensitivity to work function is greater for two-photon as compared with three-photon processes.  
The DC field variation and intensity distribution at the surface of the tip can be determined for a given tip geometry by simulation, and for geometries close to the ones simulated here, could be estimated from Fig.~\ref{fig:FieldEnhancement}.  In practice, deviations of the tip shape from the idealized hemisphere assumed here could also introduce some modifications of the emission pattern.

In the model developed above for the amplitude and angular distribution of photoemission, the values of the work function over the surface of the tip and the relative probabilities of absorbing $n$ photons are left as parameters that must be determined experimentally.  The model is sufficiently general that a single measurement of the relevant work function map suffices to predict the angular distributions for an $n$-photon process at any excitation wavelength.  For experiments using a single excitation wavelength, a measurement of the relative probabilities of photon absorption at a specific pulse duration and intensity can be generalized to predict the total tip current and relative contributions of different photon orders for arbitrary laser pulse parameters.

\section{Experimental Methods\label{Sec:Methods}}
We measure the laser-induced emission pattern and total current from a tungsten nanotip and use this to infer a work function map and relative two- and three-photon excitation probabilities under illumination with a titanium:sapphire laser.

A polycrystalline tungsten nanotip is mounted in ultrahigh vacuum, roughly 1.6~cm from a dual-chevron microchannel plate and phosphor screen assembly, which together serve as an amplified, spatially resolved electron detector.  Pulses from a Ti:sapphire oscillator (center wavelength 790~nm, repetition rate 81~MHz) with polarization parallel to the axis of the tip are focused onto the tip by a spherical mirror ($f=4.5$~mm).  The pulse intensity full width at half maximum (FWHM) is estimated to be 30 fs through an interferometric autocorrelation trace recorded using three-photon electron emission from the tip as a nonlinear element.  The focused spot size is estimated to be 2.5~$\mu$m, taking into account spherical aberration of the mirror.

Immediately prior to data collection, the tip is cleaned through field evaporation in a $2\times 10^{-6}$~torr helium atmosphere.  All measurements are taken at room temperature and pressures at or below $5\times 10^{-10}$~torr.  Measurements of total photocurrent are recorded by chopping the laser at 81~Hz and measuring the current at the grounded front face of the microchannel plate with a current-to-voltage preamplifier followed by a lock-in amplifier.  Spatially resolved measurements of photocurrent are recorded by imaging the phosphor screen with a camera outside the vacuum chamber.  The absolute number of electrons per pulse corresponding to a given lock-in signal was calibrated by visually counting single-electron detection events in a low-signal regime.

\section{Results and discussion}\label{Sec:Results}
Based on measured field emission and field ion microscope images, emission comes from a single (110)-oriented crystal at the tip apex, which is typical for polycrystalline tungsten tips \cite{Daniel1942}.  The pattern shows clear spots corresponding to the (310) and (111) directions, which have work functions of 4.32 and 4.45 eV, respectively \cite{Kawano2022}.

\begin{figure}%
\includegraphics[width=8.6cm]{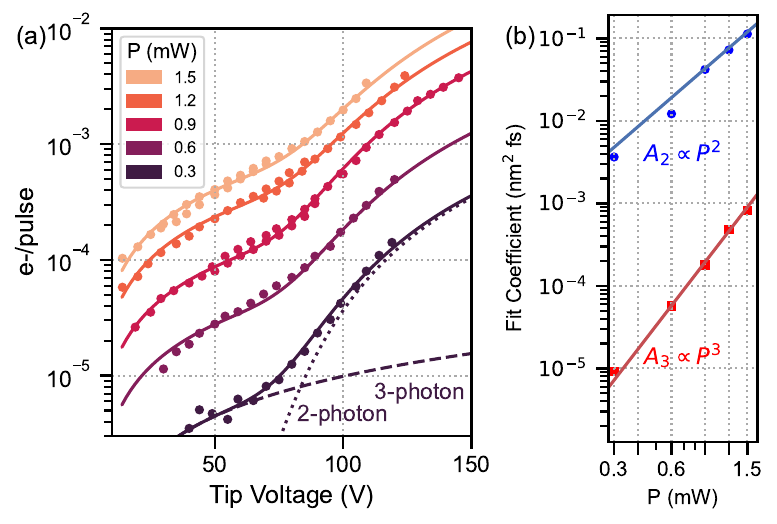}
\caption{\label{fig:ofn_fit2} Laser-induced current vs.~tip voltage for 0.3, 0.6, 0.9, 1.2, and 1.5 mW average laser power.  (a) Curves fit to the model $A_2 J_{2\text{,r}}(F, T) + A_3 J_{3\text{,r}}(F, T)$.  Field $F=V/kr$ with $kr=69$~nm.  The photon energy is $\hbar \omega = 1.575$~eV.  $A_2$, $A_3$ are fit parameters as are work functions for the two- and three-photon components (fitted values are $\phi_2 = 4.61$~eV and $\phi_3 = 4.87$~eV). Dotted and dashed lines show functional form of Eq.~\ref{eq:Jnrescaled} for two- and three-photon processes, respectively.  (b) Log-log plot of the fit coefficients, $A_2$ and $A_3$, as a function of laser power.  Fits to second and third order power laws are shown.}%
\end{figure}

Figure \ref{fig:ofn_fit2} shows the dependence of laser-induced current on tip voltage for five values of laser power ($P = 0.3$, 0.6, 0.9, 1.2, and 1.5~$\mu$W).  The data were calibrated as described in the previous section and corrected for the detector efficiency, which varies from 0.5--0.6 over the range of measured voltages.  
We define a rescaled current density,
\begin{equation}
J_{n\text{,r}} = \int_0^\infty N_n (W, T)\, D(W, F)\; dW,
\label{eq:Jnrescaled}
\end{equation}
which captures the field-dependence of the current and can be computed numerically using equations \ref{eq:TransmissionFunction} and \ref{eq:SupplyFunction}.  The datasets are fit to a sum of contributions from two- and three-photon processes, $A_2 J_{2\text{,r}}(F, T) + A_3 J_{3\text{,r}}(F, T)$.

The relationship between the tip voltage and applied field depends on the details of the tip geometry; it can be expressed as $F = V/kr$, where $r$ is the radius of the tip, and $k$ is a dimensionless geometrical parameter, usually around 5.  The value of $kr$ was determined by measuring the DC field emission $I$-$V$ curve and fitting with the Fowler-Nordheim equation.  From the resulting value $kr=69$~nm, we infer a tip radius of $\sim12.5$~nm using a finite-element model of the fields with a tip opening angle of $3.5^\circ$ (as measured with SEM) and a tip to screen distance of 1.6~cm.

The theory curves were computed for a temperature of 300~K.  Since the measured current is an average over emission from the entire tip, which includes regions of different work function, we fit an effective work function which best describes the behavior.  However, since three-photon emission is less sensitive to work function, a larger range of work functions contributes significantly to the three-photon current compared with the two-photon current (discussed in more detail below).  This is accommodated by allowing the effective work function for the two- and three-photon processes to differ, which leads to $\phi_2 = 4.60$~eV and $\phi_3 = 4.87$~eV.  Note that we required $\phi_2$ and $\phi_3$ to be the same for all datasets although in reality there is a small drift in work function over the duration of data collection due to some accumulation of adsorbates on the tip.

The fitted coefficients $A_2$ and $A_3$ giving the relative amplitudes of the two- and three-photon processes scale with $P^2$ and $P^3$ respectively, as expected.

\begin{figure}%
\includegraphics[width=\columnwidth]{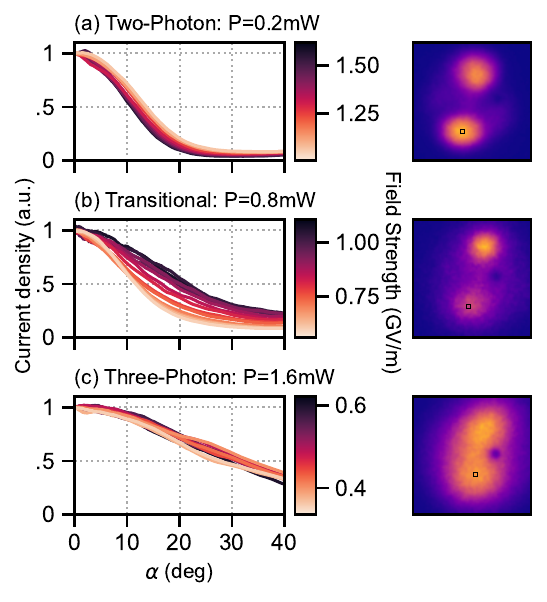}
\caption{\label{fig:emissionProfiles} Electron emission profile in different emission regimes.  Left side: current density vs.~emission angle $\alpha$.  Curves are normalized to the value at $0^\circ$.  (a) Two-photon emission results in a narrow emission pattern, which broadens slightly as voltage increases.  (b) Transition between three-photon and two-photon regime.  As the tip voltage increases, emission transitions from majority three-photon to majority two-photon, and the angular distribution narrows (c) three-photon emission results in a broad emission pattern.  Right side: example emission patterns in the different regimes, with (310) direction marked: (a) bias field: 1.6~GV/m, laser power: 0.2~mW (b) 0.85 GV/m, 0.8~mW (c) 0.65 GV/m, 1.6~mW.  The dark spot to the right of the center in the images is a damaged spot on the detector.}
\end{figure}

For all of the values of laser power measured, at sufficiently high tip voltage, the emission is dominated by two-photon emission.  Using the theoretical model, in this regime, the amount of current that is due to over-the-barrier processes is of the same order of magnitude as current due to tunneling.  When two-photon emission dominates, the emission pattern is more tightly localized around the bright (310) spots.  This is analyzed quantitatively in Fig.~\ref{fig:emissionProfiles}.

Typical emission patterns in the two- and three-photon regimes are shown in Fig.~\ref{fig:emissionProfiles}.  These are measured with the same tip as before, but after substantial field evaporation so that the tip radius has increased from 12.5 to $\sim15$~nm.  The bright spots correspond to the (310) and (130) directions (located symmetrically with respect to the (110) direction at the tip apex, which appears dark).  Note that for most electron optics applications, (310)- or (111)-oriented single-crystal tungsten tips are utilized to ensure a strong emission spot on-axis.

To compute the angular emission distribution, the center of the emission spot is estimated, and the image intensity is interpolated along circles of different $\alpha$ (emission angle measured from the (310) direction) and averaged to obtain a current density value.  Figure \ref{fig:emissionProfiles}(a)--(c) shows emission profiles computed using the lower emission spot.  The screen position to emission angle calibration was obtained by simulating the electron trajectories from the tip to the screen; the assumed distance from tip to screen was adjusted slightly to obtain the correct distance between the (310) and (130) spots on the detector.  

In the three-photon regime, the emission pattern is broader, with a half-width at half-max of $\approx 32^\circ$.  In the transition regime, where both two- and three-photon processes contribute, the emission pattern gradually becomes narrower as the tip voltage is increased and two-photon emission becomes more important.  In the two-photon regime, the emission width becomes much narrower, in the range $\sim 11$--$14^\circ$; it increases slightly as the voltage is increased.  It follows that for fixed laser intensity, there is an optimum applied field, near the field threshold where two-photon emission becomes dominant, where the angular divergence of the beam is minimized. 

\begin{figure}%
\includegraphics[width=\columnwidth]{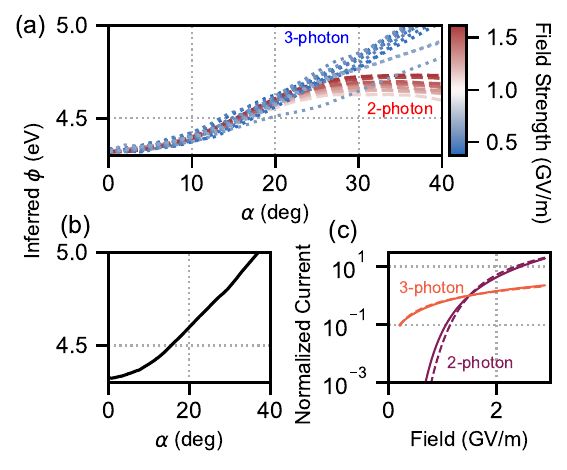}
\caption{\label{fig:phiMaps} (a) Work function map computed as a function of $\alpha$, assuming $\phi = 4.32$~eV in the (310) direction.  Each of the curves from Fig.~\ref{fig:ofn_fit2}(a) and (c) has been converted to an inferred work function using the theoretical model of Section \ref{subsec:MultiphotonTheory}.  The color scale is chosen so that lighter colored curves are closer to the transition region for both two- and three-photon emission. (b) Work function map obtained by averaging the three-photon emission work function maps in (a).  (c) Solid lines: predicted two- and three-photon emission as a function of bias field, calculated by integrating the theoretical emission computed for the measured work function distribution up to an angle of $60^\circ$.  Dashed lines: theoretical current for effective work functions $\phi_2 = 4.6$~eV and $\phi_3 = 4.87$~eV.  Curves have been normalized to their value at 1.5~GV/m.}
\end{figure}

Using the emission profiles, we can use the theoretical emission curves to infer a map of the average work function as a function of angular coordinate on the tip apex.  Here, we make the approximation that the work function has cylindrical symmetry around the (310) direction and neglect variations in the DC and optical field strengths.  Figure \ref{fig:phiMaps} shows work function maps for two- and three-photon emission, computed assuming a work function of 4.32~eV in the (310) direction.  At angles above $\sim 25$ degrees, they diverge.  For these larger angles, the current density in the two-photon case becomes small, and the work function inference is much more sensitive to systematic shifts coming from imperfect background subtraction.  Three-photon emission is also less sensitive to variations in the DC electric field (see Fig.~\ref{fig:WorkFunction_FieldVariation}), making the three-photon work function map more accurate.

Using the average of the work function maps determined using three-photon emission, it is possible to make a better theoretical prediction of the emitted two- and three-photon emission currents, integrated over the varying work function of the tip.  This procedure results in curves that are approximately the same shape as the curves of effective work functions $\phi_2$ and $\phi_3$ found via fitting in Fig.~\ref{fig:ofn_fit2} and confirms the validity of the use of effective work functions in the original fits.

The relative contributions of two- and three-photon emission, as well as the total current from the tip, can be predicted for a wide range of laser parameters beyond those studied here.  The total current from the tip is proportional to the integral of $J_{n\text{,r}}$ over the surface of the tip:
\begin{equation}
\int J_{n\text{,r}} (\phi, F) dA \approx J_{n\text{,r}}(\phi_\text{eff}, F) A_\text{eff},
\label{eq:EffectiveArea}
\end{equation}
which defines an effective emission area $A_\text{eff}$.  Comparing this with an expression for the number of electrons per pulse, it follows that the experimentally determined fit coefficients $A_n$ are related to the original parameters $a_n$ via
\begin{equation}
A_n = a_n A_\text{eff} \int I^n(t) dt,
\label{eq:An}
\end{equation}
which permits scaling to laser pulses with different temporal intensity profiles $I(t)$.
\section{Conclusions}

We present a numerical model for nanotip photoemission and experimentally characterize nanotip photoemission in the two- and three-photon emission regimes as a function of applied voltage.  We measure both the total current emitted and the changes in emission pattern that occur as a function of voltage and laser power.

The numerical model agrees well with the measured data, and the experimentally determined parameters $A_2$ and $A_3$ enable quantitative predictions of how emission scales under different experimental conditions.  Combining the numerical model with the emission pattern data, we can infer a work function map in the vicinity of the (310) direction.  This work function map can be used with simulated models of the DC and optical field distributions at the tip surface to make detailed, quantitative predictions for the angular divergence of the beam.

For most applications, it is desirable to have as many electrons in the pulse as possible to improve signal-to-noise, provided this does not unacceptably degrade the spatial or temporal resolution of the experiment.  In the measurements reported here, the count rates are always much less than one electron per pulse to avoid outgassing of the electron detector.  However, up to at least $\sim1000$ electrons per pulse have been reported in experiments on nanotips using 30~fs pulses \cite{Bormann2010} and sub-5-fs pulses \cite{Schoetz2021}.  Nanotip arrays are being actively studied for experiments requiring higher numbers of electrons per pulse without sacrificing time resolution \cite{Brueckner2024}.

Working with a single tip, the total number of electrons in the pulse should scale with emission area, which should increase approximately as the square of the tip radius.  Accepting a somewhat longer electron pulse duration can also be a strategy for increasing the number of electrons in the pulse.  For example, in the 1--10~keV range suitable for studying 2D materials \cite{Badali2016}, kinematics typically lead to a broadening of the pulse to the order of $\sim 1$~ps by the time it reaches the sample.  Thus the laser pulse duration and number of electrons in the pulse could be substantially increased without impacting the time-resolution of such an experiment.

Ultrafast electron optics opens up many exciting new opportunities for research and development.  Achieving the potential of this technology, however, requires new design approaches to overcome additional constraints compared with conventional electron optics.  Some of these constraints arise from the need to consider broadening of the pulse in the optical system due to the time delays of different electron trajectories.  Additionally, for large numbers of electrons in a pulse, the typical electron-electron spacing may be much smaller than in a conventional instrument, leading to significant effects from Coulomb repulsion on both the longitudinal and transverse properties of the pulse.  The model and empirically-measured parameters presented here enable accurate simulation and design and thus will facilitate realization of novel instruments based on ultrafast nanotip electron sources.

\appendix
\section{Fits to photoemission curves}
\begin{figure}[h]
\includegraphics[width=\columnwidth]{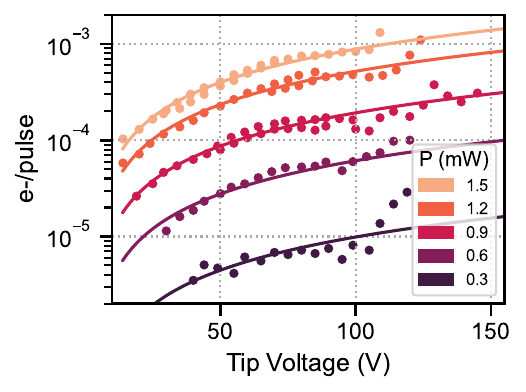} 
\caption{\label{supfig:ofn_fit2} Fit to the three-photon component.  Same data and fits as Fig.~\ref{fig:ofn_fit2}(a) of the main text; the two-photon component has been subtracted from each dataset and the three-photon fit is shown.}%
\end{figure}

When implementing fits to the $I$-$V$ data in Fig.~\ref{fig:ofn_fit2}, we found that a sensitive test of the quality of the fit is to make a ``residual" plot of the original data with the fitted two-photon component subtracted.  This is shown in Fig.~\ref{supfig:ofn_fit2}.  The residuals represent the three-photon contribution to the emission.  The solid curve shows the three-photon component of the fit.

\section{Analytic expressions for photoemission}
In some parameter regimes, the current is dominated by over-the-barrier emission, in which case, making the approximation $D\approx 1$ above the barrier and zero below it, Eq.~\ref{eq:MultiphotonComponent} can be integrated analytically to yield \cite{Girardeau-Montaut1995}
\begin{align}
J_n =& \; a_n(I_L(t))^n \frac{e m (k_B T)^2}{2\pi^2\hbar^3} \nonumber \\
&\times \text{Li}_2(-\exp(\large[\phi-n\hbar\omega-\sqrt{\frac{e^3F}{4\pi\epsilon_0}}\large]/k_B T)),
\label{eq:JnAnalytic}
\end{align}
where $\text{Li}_2$ is the polylogarithm of order 2.  A useful approximation (even at room temperature) is to take $T\rightarrow 0$, where the previous equation simplifies to
\begin{equation}
J_n = a_n(I_L(t))^n \frac{e m }{4\pi^2\hbar^3} \left(\phi - n\hbar\omega - \sqrt{\frac{e^3F}{4\pi\epsilon_0}}\right)^2.
\end{equation}
In these equations, $\sqrt{W/(W-n\hbar\omega)}$ has been approximated as a constant and folded into $a_n$ on the assumption that it varies slowly over the region where the integrands are large.

In a regime where over-the-barrier emission is negligible, an analytic equation for the current can be obtained by using the Fowler-Nordheim equation, with a work function reduced by $n$ times the photon energy.  However, in the range of laser parameters used here, the emission is never dominated by photo-assisted field emission---it either encompasses significant contributions from both over-the-barrier and photo-assisted field emission or is dominated by over-the-barrier emission.

\section{Finite element model of optical field enhancement}
Optical fields were simulated by finite element analysis using an axisymmetric partial wave expansion (up to 6 partial waves included) of the incident plane wave \cite{comsol60}.  For the complex index of refraction of tungsten at 800 nm, we used $3.56+2.76i$ \cite{Palik1997}.  The polarization component normal to the surface is evaluated just outside the tip (at a distance of 1.5\% of the tip radius) to avoid numerical artifacts from interpolation across a boundary where the fields are discontinuous.  The simulation was validated by comparing the maximum optical field enhancement to values published in \cite{Thomas2015} for several tip geometries.  

\begin{acknowledgments}
We thank Jason Mativi and Michael Taylor of the Williams Science Center Machine Shop for help constructing the apparatus, and Nancy Piatczyc for assistance with electron microscopy.  We also thank Emily Stump, Ellery Galvin, Michael Arena, Abdullah Nasir, Heather Kurtz, Joshua Reynolds, Ilana Albert, Declan Daly, Benny Weng, Katya Ulyanov, Katie Brockmeyer, and Da-Yeon Koh for help with aspects of the experimental apparatus, and Kate Jensen and Seth Foreman for helpful discussions.
This work was supported by a Cottrell Scholar Award from the Research Corporation for Science Advancement.
\end{acknowledgments}

\bibliography{../../../../eguide2}

\end{document}